\documentclass[letterpaper, 10pt, conference]{ieeeconf}
\IEEEoverridecommandlockouts  % to override locked commands

\renewcommand{\baselinestretch}{0.98}
\usepackage{cite}
\usepackage{amsmath,amssymb,amsfonts}
\usepackage{algorithmic}
\usepackage{graphicx}
\usepackage{textcomp}
\usepackage{hyperref}
\usepackage{xcolor}
\usepackage[ruled,vlined]{algorithm2e}
\usepackage[capitalize]{cleveref}
\usepackage{tikz}
\usetikzlibrary{patterns}

\def\BibTeX{{\rm B\kern-.05em{\sc i\kern-.025em b}\kern-.08em
    T\kern-.1667em\lower.7ex\hbox{E}\kern-.125emX}}

\title{\LARGE \bf
Maximal Compatibility Matching for \\Preference-Aware Ride-Hailing Systems

}
\author{Avalpreet Singh Brar$^{1,2}$,  Rong Su$^{2}$, Jaskaranveer Kaur$^{1}$, Xinling Li$^{3}$, Gioele Zardini$^{3}$   % <-this % stops a space
\thanks{This study is supported under the RIE2020 Industry Alignment Fund – Industry Collaboration Projects (IAF-ICP) Funding Initiative, as well as cash and in-kind contribution from the industry partner(s).}%
\thanks{$^{1}$Continental Automotive, Singapore ({ avalpreet.singh.brar@continental-corporation.com, jaskaranveer.kaur@continental-corporation.com}).}
\thanks{$^{2}$School of Electrical and Electronic Engineering, Nanyang Technological University, Singapore. (brar0002@e.ntu.edu.sg, rsu@ntu.edu.sg).}
\thanks{$^{3}$Laboratory for Information and Decision Systems, Massachusetts Institute of Technology, Cambridge, MA, USA. (xinli831@mit.edu, gzardini@mit.edu).}}
\begin{document}

\maketitle

\begin{abstract}
This paper presents the Maximal Compatibility Matching (MCM) framework, a novel assignment strategy for ride-hailing systems that explicitly incorporates passenger comfort into the matching process. 
Traditional assignment methods prioritize spatial efficiency, but often overlook behavioral alignment between passengers and drivers, which can significantly impact user satisfaction. 
MCM addresses this gap by learning personalized passenger comfort zones using gradient-boosted decision tree classifiers trained on labeled ride data, and by modeling driver behavior through empirical operating profiles constructed from time-series driving features. 
Compatibility between a passenger and a driver is computed as the closed-form volume of intersection between their respective feature-space regions. 
These compatibility scores are integrated into a utility-based matching algorithm that balances comfort and proximity through a tunable trade-off parameter. 
We validate the framework using a Unity-based driving simulator with real-time passenger feedback, demonstrating that MCM enables more personalized and socially acceptable matchings while maintaining high levels of operational performance.
\end{abstract}

\section{Introduction}
In recent years, ride-hailing services such as Uber, Lyft, and DiDi have experienced rapid and sustained market growth. With a projected compound annual growth rate exceeding 8.75\%~\cite{liu2022machine}, the global market is expected to reach \$230B by 2026~\cite{intelligence2021epoxy}. These platforms offer a flexible alternative to private vehicle ownership by eliminating the burdens of maintenance and parking while enabling personalized, on-demand mobility. However, this flexibility introduces significant operational complexity, particularly in managing dynamic, real-time requests across large urban fleets~\cite{ZardiniAnnRev2022}.

Recent literature has identified several key challenges that impact the operational efficiency of ride-hailing systems, including \emph{supply positioning}~\cite{brar2021dynamic, wallar2018vehicle}, \emph{fairness} in passenger allocation ~\cite{guo2023fairness,brar2020ensuring}, \emph{platform incentive} strategies~\cite{sadeghi2019re, brar2024integrated}, and  \emph{adherence} to platform recommendations~\cite{brar2024vehiclerebalancingadherenceuncertainty}. Central to all of these issues is the \emph{vehicle-passenger matching} problem. Upon receiving travel requests, platforms must rapidly assign available vehicles to passengers in a manner that optimizes critical performance metrics such as waiting time, service rate, and revenue.
This problem is commonly modeled as a weighted bipartite matching problem~\cite{cattrysse1992survey}, often solved via the Hungarian algorithm~\cite{kuhn1955hungarian} to maximize the total matching utility.
In typical formulations, the edge weights reflect revenue potential or travel distance~\cite{yan2020dynamic,feng2024two,guo2021robust,ashlagi2018maximum}, and similar criteria are used in learning-based approaches~\cite{liu2022deep,ke2019optimizing,zhou2019multi}.
While these strategies improve operational efficiency and profitability, they overlook behavioral heterogeneity, that is, the variation in passenger travel preferences and driver behavior.
These qualitative factors strongly influence user experience~\cite{ben1985discrete,ben2002hybrid}.
To ensure passenger comfort and satisfaction, future matching algorithms must integrate both economic and experiential considerations into the decision-making process.

%In the context of matching for ride-hailing systems, the weights of the bipartite graph are usually calculated based on the revenue of serving the passenger by the vehicle or the distance between the vehicle and the passenger~\cite{yan2020dynamic,feng2024two,guo2021robust,ashlagi2018maximum}. Beyond optimization-based matching, revenue and cost are also common choices in studies on learning-based matching, where the rewards are often defined by the monetary benefit of the matching \cite{liu2022deep,ke2019optimizing,zhou2019multi}. While this weight design aims to improve the service profitability, it ignores the heterogeneity in passengers' travel preferences and drivers' driving behavior, which significantly impacts the experience of passengers using the service. To guarantee the user experience, it is important to account for the qualitative factors that influence user comfort and satisfaction and embed them into the process of matching decision-making.

\subsection{Related Literature}
To address the limitations of matching strategies based solely on operational objectives, recent research has explored compatibility-based matching mechanisms that incorporate user-centric preferences into the decision-making process.
These approaches augment traditional optimization goals, such as minimizing distance or maximizing revenue, with social compatibility scores that reflect both passenger and driver preferences, often derived from personal or behavioral attributes.
For instance~\cite{aydin2020matching} introduced a ``joint socialnness score'' based on factors such as gender, age group, employment affiliation, and openness to social interaction.
This score was integrated into the matching graph as an additional weighting metric, enabling the system to prioritize socially agreeable matches.
Their findings demonstrated that even minor compromises in travel efficiency led to significant improvements in social compatibility.
A similar approach was taken in~\cite{zhang2018mobility} in the context of carpooling, where compatibility between co-riders was definend through stated preferences.
Their results showed that incorporating such preferences significantly improved user satisfaction.
Building on this work,~\cite{dastani2024user} proposed a multi-attribute compatibility model that formalizes passenger preferences into a structured matching algorithm.
This model integrates both static attributes (e.g., demographics, vehicle type) and dynamic behavioral factors (e.g., driving style) to compute a mutual compatibility score for each passenger-vehicle pair.
In parallel, another body of work has focused on empirical and data-driven methods to estimate compatibility. Cui et al.~\cite{cui2021understanding} derived a probabilistic compatibility index from large-scale user surveys on shared mobility preferences. 
This index was shown to be predictive of ride satisfaction and was used as a guiding metric for assignment decisions. Complementarily, Fajardo-Delgado et al.~\cite{fajardo2022stable} modeled the matching process using stable matching theory, where utility and reputation scores were dynamically updated to ensure that each match was mutually optimal under a set of individual constraints.

Collectively, these studies demonstrate that compatibility is both quantifiable and operationalizable.
Specifically, they underscore the growing recognition that accounting for passenger-driver preferences can lead to more socially acceptable, comfortable, and ultimately sustainable mobility services.

\subsection{Statement of Contribution}
This paper presents the MCM framework, a new assignment strategy for ride-hailing systems that integrates passenger comfort with spatial efficiency.
MCM combines data-driven modeling of passenger comfort zones with empirical characterization of driver operating behavior, yielding a closed-form compatibility score based on feature-space overlap.
The framework formulates passenger-driver assignment as a utility-maximizing optimization that balances comfort and proximity through a tunable trade-off.
We validate the approach through a simulator-based study with real-time human feedback, demonstrating its effectiveness in capturing preference-aware matchings while preserving operational performance.
\subsection{Organization of Manuscript}
The remainder of this paper is organized as follows. 
\cref{sec:formulation} introduces the MCM framework, describing the modeling of passenger comfort zones and driver operating profiles, the derivation of the compatibility score, and the utility-based assignment algorithm.
\cref{sec:experiment} presents the experimental setup and simulation-based evaluation, including the feature extraction pipeline, classifier performance, and a comparative analysis of distance-based, comfort-based, and utility-based matchings.
\cref{sec:conclusion} concludes with a discussion of key findings and future research directions.

\section{Mathematical Formulation: Maximal Compatibility Matching} \label{sec:formulation}
In this section, we formally define the MCM framework that models the problem of pairing passengers and drivers based on their respective comfort and operating regions in a \(N\)-dimensional feature space. 
The process consists of four sequential stages.

\subsection{Passenger Comfort Zone}
Each passenger \(i \in \mathcal{P}\) provides labeled feedback on ride experiences, indicating whether a feature vector \(\mathbf{x} \in \mathbb{R}^N\) is perceived as comfortable (\(y=0\)) or uncomfortable (\(y=1\)).
These features may include driving behavior, driver's environment, and other ride attributes.
The objective is to learn a non-linear classification function \(G_i(\mathbf{x}) : \mathbb{R}^N \rightarrow \{0, 1\}\) that captures the passenger's individual comfort preferences.
To approximate the map, we train a gradient-boosted decision tree classifier, which builds an ensemble of decision trees through iterative residual fitting.
The process of generating the decision boundary involves the following steps:

\paragraph{Step 1: Input Space and Label Definition}
Let \(\mathbf{x} = (x_1, x_2, \dots, x_N)^\top \in \mathbb{R}^N\) represent a ride configuration across \(N\) features.
Furthermore, let \(y_i(\mathbf{x}) \in \{0, 1\}\) be the passenger's feedback label \(i\), where \(1\) indicates an uncomfortable ride.
Then, the training set for the passenger \(i\) is:
\begin{equation}
\mathcal{X}_i = \{(\mathbf{x}^{(k)}, y_i^{(k)})\}_{k=1}^{K_i}
\end{equation}

\paragraph{Step 2: Tree-Based Partitioning of the Feature Space}

Each decision tree in gradient boosted tree based classifier recursively partitions the feature space using axis-aligned splits of the form:
\begin{equation}
x_d < \theta_d
\end{equation}
where \(d \in \{1, \dots, N\}\) is the selected feature and \(\theta_d\) is a threshold chosen to minimize classification error or loss. These splits define hyper-rectangular leaf regions \(\{R_m\}\) in which each region is assigned a probability estimate \(p_m\) for \(P(y=1 \mid \mathbf{x} \in R_m)\).

\paragraph{Step 3: Boosted Ensemble Construction}

The gradient boosted tree based classifier combines \(T\) such decision trees into a strong classifier:
\begin{equation}
G_i(\mathbf{x}) = \sigma\left( \sum_{t=1}^{T} g_t(\mathbf{x}) \right),   
\end{equation}
where:
\begin{itemize}
    \item Each \(g_t \in \mathcal{F}\) is a tree-based function,
    \item \(\sigma(z) = \frac{1}{1 + e^{-z}}\) is the sigmoid function,
    \item The model is trained to minimize the regularized objective:
    \begin{equation}
    \mathcal{L} = \sum_{k=1}^{K_i} \ell(y_i^{(k)}, \hat{y}^{(k)}) + \sum_{t=1}^{T} \Omega(g_t),    
    \end{equation}
    with \(\ell\) being the logistic loss and \(\Omega\) being a regularization term to control model complexity.
\end{itemize}

\paragraph{Step 4: Implicit Decision Boundary Formation}

\begin{figure}[t]
    \centering
    \begin{tikzpicture}[scale=1.2]

        % Draw G as a downward staircase (step function)
        \draw[black, semithick]
            (0,4) -- (0.5,4) -- (0.5,3) -- (2,3) -- (2,2) -- (3,2 ) -- (3,1) -- (4,1) -- (4,0) -- (0,0) -- cycle;
        \node at (2.0,0.4) {\scriptsize $G_i(\mathbf{x}): \text{Comfort Zone of Passenger \(i \in \mathcal{P}\)}$};

        % Dotted horizontal lines for step levels
        \draw[dotted] (0,3) -- (0.5,3);
        \draw[dotted] (0,2) -- (2,2);
        \draw[dotted] (0,1) -- (3,1);

        % Draw red box f_j
        \draw[fill=red!20, opacity=0.6] (1.5,1.5) rectangle (4.5,3.5);
        \node at (4.2, 3.7) {\scriptsize $f_j:\text{Operating Zone of Driver \(j \in \mathcal{D}\)}$};

        % Draw intersection volume shading
        \begin{scope}
            \clip 
                (0,4) -- (0.5,4) -- (0.5,3) -- (2,3) -- (2,2) -- (3,2) -- (3,1) -- (4,1) -- (4,0) -- (0,0) -- cycle;
            \fill[pattern=north west lines, pattern color=red] (1.5,1.5) rectangle (4.5,3.5);
        \end{scope}
        \node at (2, 1.3) {\scriptsize \textcolor{red}{$A_{ij}$}};

        % Axes
        \draw[thick, ->] (-0.5,0) -- (6,0) node[right] {\small \(x_1\)};
        \draw[thick, ->] (0,-0.5) -- (0,5) node[above] {\small \(x_2\)};

    \end{tikzpicture}
    \caption{Geometric illustration of compatibility in a two-dimensional feature space.  
    The passenger's comfort zone \( G_i(\mathbf{x}) \) is depicted as a downward staircase-shaped region, while the driver's operating zone \( f_j \) is shown as a red rectangle.  
    The shaded intersection area \( A_{ij} \), highlighted with a hatched pattern, represents the region of mutual compatibility between passenger \( i \) and driver \( j \).}

    \label{fig:intersection_zone}
\end{figure}
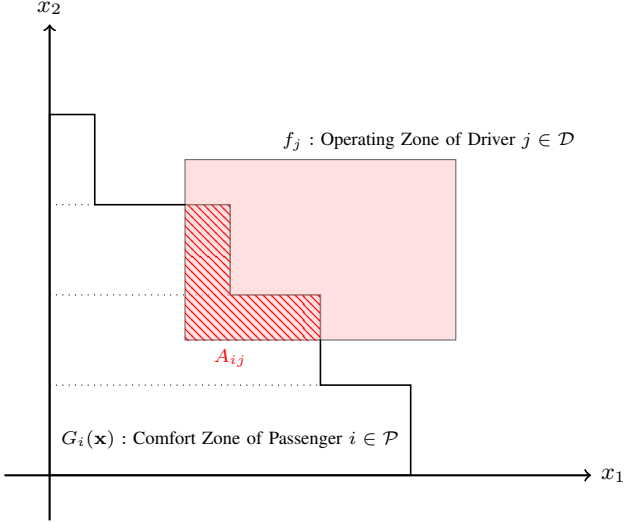

The decision boundary corresponds to the set of points in the feature space where the predicted probability of the classifier is equal to the threshold:
\begin{equation}
\partial G_i = \left\{ \mathbf{x} \in \mathbb{R}^N \;\middle|\; G_i(\mathbf{x}) = \varepsilon_i \right\}. 
\end{equation}
Here, \(\varepsilon_i \in (0,1)\) denotes the comfort threshold specified by passenger \(i\), representing the maximum score at which a ride configuration \(\mathbf{x}\) is still considered comfortable.
Owinng to the ensemble of decision trees with axis-aligned splits, the resulting decision boundary is piecewise linear (see \cref{fig:intersection_zone}), allowing the model to capture complex, non-convex comfort zones.

\paragraph{Step 5: Final Classifier Output}

The final classifier is defined as:
\begin{equation}
G_i(\mathbf{x}) = 
\begin{cases}
1 & \text{if } \sigma\left( \sum_{t=1}^{T} g_t(\mathbf{x}) \right) \geq \varepsilon_i \\
0 & \text{otherwise}.
\end{cases}    
\end{equation}  
Hence, \(G_i(\mathbf{x}) = 0\) indicates that the ride configuration \(\mathbf{x}\) lies within the passenger’s personalized \textit{comfort zone}.

\subsection{Driver Operating Zone}
Each driver \(j \in \mathcal{D}\) is associated with a history of driving behavior represented in an \(N\)-dimensional feature space. These features may include dynamic driving variables such as \textit{jerk}, \textit{acceleration}, \textit{speed}, and other context-dependent indicators relevant to ride comfort and safety.

To characterize a driver's \textit{operating zone}, we extract the empirical range of these features from previous trips. 
Specifically, for driver~$j$, we collect a set of feature vectors \(\{ \mathbf{x}_i^{(j)} \}_{i=1}^{T_j}\), where each  \(\mathbf{x}_i^{(j)} \in \mathbb{R}^N\) corresponds to a segment or instance from a recorded trip.
This collection forms an empirical distribution in the \(N\)-dimensional feature space.

To summarize the typical driving behavior in a compact and interpretable way, we construct an \emph{axis-aligned hyperrectangle} using quantile statistics, typically the 5th and 95th percentiles, for each dimension.
This hyperrectangle, representing the driver's operating envelope, is defined as:
\begin{equation}\label{eqn:quantile}
f_j = \prod_{d=1}^{N} [a_d^{(j)}, b_d^{(j)}],   
\end{equation}
where \(a_d^{(j)}\) and \(b_d^{(j)}\) denote the lower and upper quantiles (e.g., 5th and 95th percentiles) for feature~$d$ in driver~$j$'s empirical distribution.
This representation captures the core behavioral tendencies while excluding extreme outliers.
\cref{fig:intersection_zone} illustrates this quantile-based operating zone in a reduced feature space, where each axis corresponds to a distinct driving characteristic.
The red box outlines the driver's typical operating region, and its overlap with the passenger comfort zone~$G_i$ is also depicted.

\subsection{Passenger-Driver Compatibility}
The comfort region of passenger \(i \in \mathcal{P}\), denoted 
\(G_i(\mathbf{x})\), is defined in the same \(N\)-dimensional feature space.
This region is approximated ba a union of \(K_i\) axis-aligned hyperrectangles \(G_i^{(k)}\), such that:

\begin{equation} \label{eqn:classifier}
G_i(\mathbf{x}) = \bigcup_{k=1}^{K_i} G_i^{(k)}(\mathbf{x})
\quad \text{where} \quad
G_i^{(k)}(\mathbf{x}) = \prod_{d=1}^{N} [\alpha_d^{(k)}, \beta_d^{(k)}].    
\end{equation}
The compatibility between passenger \(i\) and driver \(j\) is quantified by the total volume of the intersection between the driver's operating zone \(f_j\) and the passenger’s comfort region \(G_i(\mathbf{x})\).
As all sets are axis-aligned boxes, the intersection volume between any two hyperrectangles can be computed in closed form:
\begin{equation}\label{eqn:compatibility}
A_{ij} = \sum_{k=1}^{K_i} \prod_{d=1}^{N} 
\max\left(0, \min(\beta_d^{(k)}, b_d^{(j)}) - \max(\alpha_d^{(k)}, a_d^{(j)}) \right).
\end{equation}
Geometrically, this intersection area~$A_{ij}$, illustrated in \cref{fig:intersection_zone}, corresponds to the striped region shared by the red operating box~$f_j$ and the passenger's comfort region  \(G_i(\mathbf{x})\).
The formula generalizes efficiently to high-dimensional spaces.
This compatibility score serves as the basis for passenger-driver matching.
A higher~$A_{ij}$ value indicates greater alignment between the driver's behavioral envelope and the passenger's comfort preferences.
In the next section, we leverage this score to build a matching algorithm that prioritizes compatibility.

\subsection{Utility-Based Assignment Optimization}

To achieve a socially optimal assignment that balances comfort and proximity, we define a utility score for each passenger-driver pair as a weighted combination of passenger-driver compatibility and distance. The resulting binary assignment problem is:

\[
\max_{X_{ij}} \sum_{i \in \mathcal{P}} \sum_{j \in \mathcal{D}} \left( \alpha A_{ij} - (1 - \alpha) D_{ij} \right) X_{ij},
\]

\textbf{Subject to:}
\begin{align*}
\sum_{j \in \mathcal{D}} X_{ij} &= 1 \quad && \forall i \in \mathcal{P}  \\
\sum_{i \in \mathcal{P}} X_{ij} &= 1 \quad && \forall j \in \mathcal{D}  \\
X_{ij} &\in \{0, 1\} \quad && \text{(binary decision variables)}
\end{align*}

Here, \( A_{ij} \in [0, 1] \) denotes the compatibility score between passenger \( i \) and driver \( j \), and \( D_{ij} \in [0, 1] \) is the normalized distance between them.

The parameter \( \alpha \in [0, 1] \) governs the trade-off:
\begin{itemize}
    \item \( \alpha = 0 \): the objective reduces to minimizing total distance.
    \item \( \alpha = 1 \): the objective reduces to maximizing total affinity.
    \item \( 0 < \alpha < 1 \): the optimizer balances both comfort and proximity.
\end{itemize}

\subsection{Maximal Compatibility Matching (MCM) Algorithm}

This section details the Maximal Compatibility Matching (MCM) algorithm, designed to perform passenger comfort-aware driver allocation. The algorithm operates in four main stages. First, a set of candidate passengers requesting pickup and a pool of idle drivers within a defined geo-spatial boundary are identified. Next, the pairwise distances between each passenger-driver pair are computed using a chosen distance metric (e.g., Euclidean distance or network-based travel distance). Concurrently, a compatibility matrix containing pre-computed scores for each passenger-driver pair is loaded. These compatibility scores can be generated offline using historical trip ratings and driver behavioral profiles. Finally, a utility-based optimization problem is solved to determine a matching that balances spatial proximity with passenger comfort preferences.

\paragraph{Computational Complexity}
The computational complexity of the Maximal Compatibility Matching (MCM) algorithm is governed by two main components: (i) computing pairwise distances and compatibility scores, and (ii) solving the assignment problem. The first step evaluates all possible passenger-driver pairs, resulting in a complexity of \( O(|\mathcal{P}||\mathcal{D}|) \). The second step is a linear sum assignment problem, solved using the Hungarian algorithm in \( O\left( \min(|\mathcal{P}|, |\mathcal{D}|)^2 \cdot \max(|\mathcal{P}|, |\mathcal{D}|) \right) \) time. Thus, the overall computational complexity of the MCM algorithm is:
\[
O\bigg(|\mathcal{P}||\mathcal{D}| + \min(|\mathcal{P}|, |\mathcal{D}|)^2 \cdot \max(|\mathcal{P}|, |\mathcal{D}|)\bigg),
\]
making it suitable for deployment in moderate-scale real-world mobility systems. 

\begin{algorithm}[h]
\caption{Maximal Compatibility Matching}
\KwIn{
    \( \mathcal{P} \): set of passengers with coordinates; \\
    \( \mathcal{D} \): set of idle drivers with coordinates; \\
    \( \alpha \in [0,1] \): comfort-sensitivity trade-off parameter; \\
    \( A \in \mathbb{R}^{|\mathcal{P}| \times |\mathcal{D}|} \): compatibility matrix.
}
\KwOut{Binary assignment matrix \( X_{ij} \in \{0,1\} \)}

\vspace{4pt}
\textbf{Stage 1: Compute Distance and Load Compatibility Scores}\;
\For{each passenger \( i \in \mathcal{P} \)}{
    \For{each driver \( j \in \mathcal{D} \)}{
        Compute spatial distance \( D_{ij} \); \\
        Load pre-computed compatibility score \( A_{ij} \);
    }
}

\vspace{4pt}
\textbf{Stage 2: Solve Comfort-Aware Assignment}\;
Solve the following optimization problem for comfort-sensitivity trade-off parameter \(\alpha\):
\[
\max_{X_{ij}} \sum_{i \in \mathcal{P}} \sum_{j \in \mathcal{D}} \left( \alpha A_{ij} - (1 - \alpha) D_{ij} \right) X_{ij}
\]
s.t. \(\sum_j X_{ij}=1, \sum_i X_{ij}=1, X_{ij} \in\{0,1\} \);

\Return{optimal matching matrix \( X_{ij} \)}
\end{algorithm}

\section{Experiment and Case Studies} \label{sec:experiment}
We evaluate the proposed framework using a Unity-based simulator featuring 5 drivers and 13 passengers.
Each driver operates the simulator while passengers assess the driving behavior in real time, labeling each session as either \emph{calm} or \emph{rash} based on their individual comfort thresholds.
The system records three driving features, speed, acceleration, and jerk, at a sampling rate of 10 samples per second, enabling fine-trained tracking of driver behavior.

\subsection{Passenger-Driver Compatibility}
A structured preprocessing pipeline is employed to transform the raw time-series data into a form suitable for analysis.
Each passenger's dataset comprises continuous readings of vehicle speed, acceleration, and jerk.
The data is segmented into uniform 10-second intervals, and for each interval, we extract a set of 21 statistical features, including the mean, median, standard deviation, minimum, maximum, and the 25th and 75th percentiles of each raw signal.
Each segment is labeled 0 (calm) or 1 (rash) based on the passenger's real-time assessment.

\begin{figure}[tb]
    \centering
    \includegraphics[width=1\linewidth]{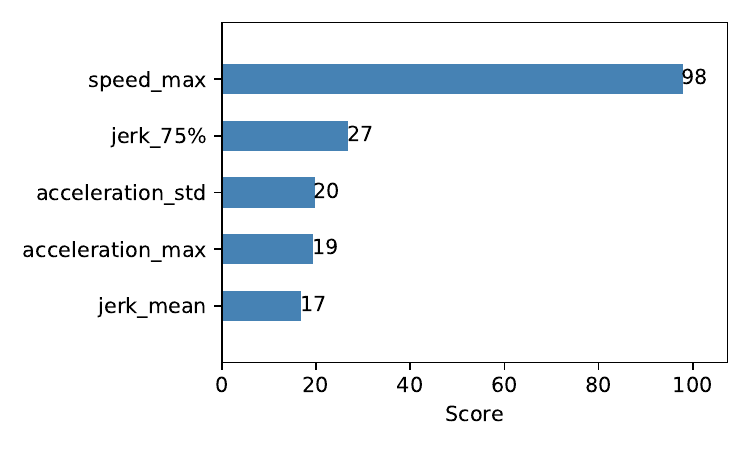}
    \caption{
    Top 5 features based on gain values computed by the gradient-boosted decision tree model. The gain metric reflects each feature's relative contribution to improving model performance, with higher values indicating greater predictive importance.
    }
    \label{fig:top_features}
\end{figure}

This feature engineering step captures critical temporal patterns in driving behavior.
For real-time deployment, the system can classify driver behavior after each interval.
For longer sessions, an aggregated \emph{urgency score}, defined as the mean of interval-level labels, is computed to summarize the overall driving style.

\begin{figure*}[tb]
    \centering
    \includegraphics[width=1\linewidth]{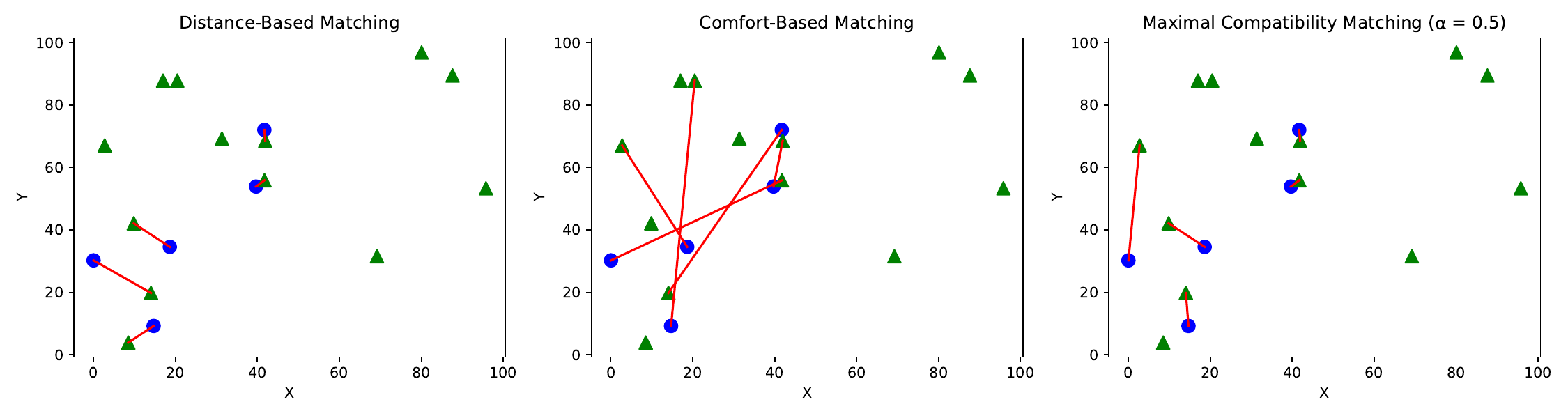}
    \caption{
    Comparison of passenger-driver matching under three assignment strategies: distance-based matching (left), which minimizes total travel distance; comfort-based matching (center), which maximizes affinity based on passenger preferences; and maximal compatibility matching (right), which optimizes a utility function combining comfort and distance with trade-off parameter $\alpha = 0.5$. The utility-based approach captures a balance between proximity and personalized satisfaction.
    }
    \label{fig:match-comparison}
\end{figure*}

\begin{table}[tb]
\centering
\caption{Classification Report} \label{Table: Classification_Report}
\renewcommand{\arraystretch}{1.4}
\begin{tabular}{|c|c|c|c|c|}
\hline
\textbf{Class} & \textbf{Precision} & \textbf{Recall} & \textbf{F1-Score} \\
\hline
0 & 1.00 & 0.91 & 0.95  \\
1 & 0.84 & 1.00 & 0.91  \\
\hline
\textbf{Accuracy} & & & 0.94 \\
\hline
\textbf{Macro average} & 0.92 & 0.96 & 0.93 \\
\hline
\textbf{Weighted average} & 0.95 & 0.94 & 0.94 \\
\hline
\end{tabular}

\end{table}
\begin{figure}[tb]
    \centering
    \includegraphics[width=1\linewidth]{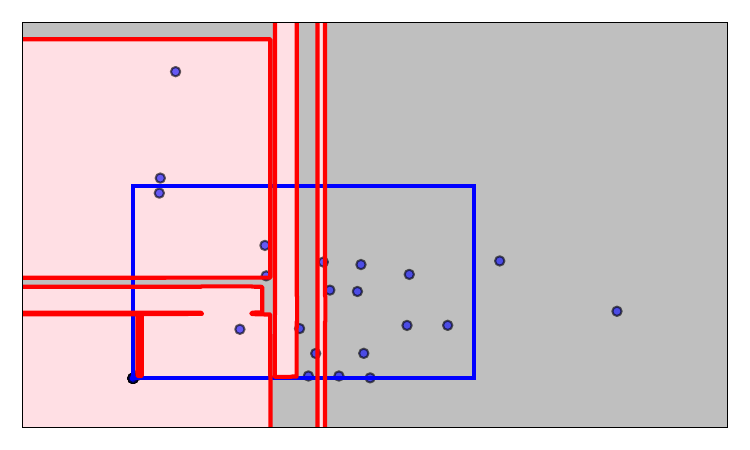}
    \caption{Visualization of passenger and driver compatibility regions in a 2D feature space. The red region with decision boundaries represents the passenger's comfort zone learned using a gradient-boosted decision tree classifier. The blue rectangle denotes the operating zone of a candidate driver. Blue scatter points correspond to the driver’s historical ride samples, used to estimate their typical operating behavior. The area of the driver operating zone that falls within the passenger comfort zone represents the passenger-driver compatibility.
    }
    \label{fig:Passenger_driver_com}
\end{figure}

A gradient-boosted decision tree classifier is used to learn from these features and perform classification. 
\cref{fig:top_features} plots the importance score of the main five features. The speed\_max feature contributes the most to the model’s predictive performance, followed by jerk\_75th percentile, acceleration\_std, acceleration\_max, and jerk\_mean. Higher gain values indicate greater influence on model decisions.
To improve label reliability, we apply a Gaussian Naive Bayes-like correction step.
For each of the 21 features, we compute the mean and standard deviation for both calm (class 0) and rash (class 1) categories.
For each data point, we calculate its likelihood under both distributions and reassign the label to the class with the higher likelihood.
This probabilistic reassignment corrects potential misclassifications caused by subjectivity or noise in real-time labeling.
Furthermore, class imbalance is evident in the dataset and adversely affects classification performance.
To mitigate this, we apply class weighting during model training to ensure that the minority class receives adequate attention.
This adjustment reduces prediction bias and improves the model's ability to generalize across both classes.
The final classification performance is summarized in \cref{Table: Classification_Report}.

As illustrated in \cref{fig:Passenger_driver_com}, the blue rectangle represents a selected driver's operating region  while red box denotes the passenger's comfort zone. This hyperrectangle is constructed by computing the lower and upper percentiles (e.g., 5th and 95th percentiles) for each of the 21 features. For visualization purposes, the figure displays only two features, capturing the dynamic interplay between driver behavior and passenger comfort. The scattered blue dots correspond to individual driving data points recorded during a trip, revealing upto what extent the driver's actions align with or deviate from the passenger's preferred comfort thresholds. The proposed framework calculates this relationship by calculating the area of intersection.

The Passenger-Driver compatibility score is defined as the volume of intersection between passenger's comfort zone and a driver's operating zone as defined in \cref{eqn:compatibility}. Given the high dimensionality of the system (21 features), Monte Carlo integration is well-suited due to its robustness to high-dimensional spaces. So, the ~$A_{ij}$ volume is being calculated using density-based Monte Carlo Sampling. 

\subsection{Matching Case Studies}
We consider three matching strategies: (i) distance-based matching, (ii) comfort-based matching, and (iii) utility-based matching. 
To illustrate the trade-offs among these approaches, \cref{fig:match-comparison} presents side-by-side visualizations of the resulting passenger-driver assignments under each formulation.
Distance-based matching prioritizes spatial proximity, yielding short travel distances but potentially neglecting individual passenger preferences.
Comfort-based matching ignores spatial constraints and instead maximizes compatibility based on passenger-defined comfort zones, often resulting in longer assignments that better align with individual preferences.
Finally, utility-based matching introduces a tunable trade-off governed by a parameter~$\alpha\in[0,1]$, interpolating between the two extremes.
At~$\alpha=0.5$, the resulting assignments strike a balance between comfort and proximity.

\begin{figure}[tb]
    \centering
    \includegraphics[width=1\linewidth]{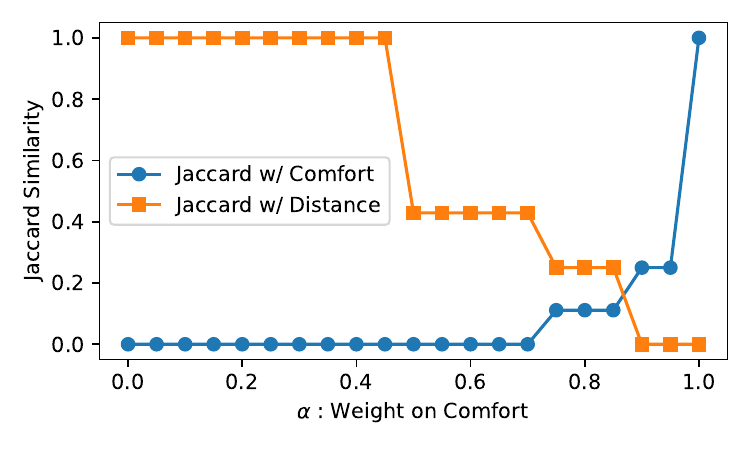}
    \caption{
    Jaccard similarity between the utility-based matching and the baseline matchings (distance-based and comfort-based) as a function of the trade-off parameter $\alpha$. As $\alpha$ increases, the utility-based matching becomes more similar to the comfort-based matching and less similar to the distance-based one, reflecting the shift in optimization priority from spatial proximity to user preference.
    }
    \label{fig:jaccard-similarity}
\end{figure}

To quantitatively compare the matchings, we compute the Jaccard similarity score, defined as:
\begin{equation}
\text{Jaccard}(M_1, M_2) = \frac{|M_1 \cap M_2|}{|M_1 \cup M_2|},    
\end{equation}
where~$M_1$ and~$M_2$ are sets of matched passenger-driver pairs.
\cref{fig:jaccard-similarity} shows how the similarity between the utility-based matching and the two baselines evolves as a function of~$\alpha$.
As expected, increasing~$\alpha$ shifts the solution from proximity-driven to preference-driven behavior: similarity with comfort-based matching increases, while similarity with distance-based matching declines.
This confirms that the utility-based formulation provides fine-grained control over matching priorities.

\section{Conclusion}
\label{sec:conclusion}
This paper introduced the MCM framework, a novel assignment strategy for ride-hailing platforms that jointly considers passenger comfort and spatial efficiency.
By learning personalized comfort zones from passenger feedback and constructing empirical driver operating profiles, MCM computes a closed-form compatibility score that quantifies behavioral alignment.
The resulting utility-based optimization formulation enables flexible assignment strategies, ranging from purely distance-based to fully preference-driven, through a tunable trade-off parameter.
Experimental results from a Unity-based simulator demonstrate that MCM can effectively capture nuanced passenger preferences and improve assignment quality without significant compromise in efficiency.
Future work will explore large-scale deployment, dynamic learning of comfort thresholds, and integration of fairness and multi-passenger scenarios into the matching process.

\bibliographystyle{IEEEtran}
\bibliography{ref.bib}

\end{document}